\documentclass[11pt]{article}

\usepackage[numbers,sort&compress]{natbib}
\usepackage{euscript}
\usepackage{amssymb}
\usepackage{amsfonts}
\usepackage{amsbsy}
\usepackage{epsfig}
\usepackage{amsthm}
\usepackage{amscd}
\usepackage{amstext}
\usepackage{verbatim}
\usepackage{amsmath}
\usepackage{cancel}
\usepackage{capt-of}
\usepackage[pdftex]{hyperref}

\usepackage{bookmark}
\usepackage{tensor}
\newcommand{\me}{\mathrm{e}}
\newcommand{\md}{\mathrm{d}}
\newcommand{\bea}{\begin{eqnarray}}
\newcommand{\eea}{\end{eqnarray}}
\newcommand{\be}{\begin{equation}}
\newcommand{\ee}{\end{equation}}

\begin{document}

\begin{center}

{ \Large {\bf
Magnetothermoelectric DC conductivities  from holography models
with hyperscaling factor in Lifshitz spacetime }}

\vspace{1cm}

Ze-Nan Chen $^{1}$,~~~ Xian-Hui Ge $^{1,2}$,~~~Shang-Yu Wu$^{3}$, ~~~Guo-Hong Yang$^{1}$,~~~Hong-Sheng Zhang $^{4}$

\vspace{1cm}

{\small
$^1${\it \small Department of Physics, Shanghai University, Shanghai 200444, P.R. China}\\
$^2${\it\small Department of Physics, University of California at San Diego, CA92093, USA}\\
$^3${\it\small Department of Electrophysics, National Chiao Tung University, Hsinchu 300, Taiwan, ROC}\\
$^4${\it\small Center for Astrophysics, Shanghai Normal University, Shanghai 200030, P.R. China}}
\vspace{1.6cm}

\end{center}

\begin{abstract}
We investigate an Einstein-Maxwell-Dilaton-Axion holographic model and obtain two classes of  a charged black hole solution with a dynamic exponent and a hyperscaling violation factor when a magnetic field presents.  The magnetothermoelectric DC conductivities are then calculated in terms of horizon data by means of holographic principle. We find that  linear  temperature dependence resistivity and quadratic temperature dependence inverse Hall angle can be achieved in our model. The well-known anomalous temperature scaling of the Nernst signal and the Seebeck coefficient of cuprate strange metals are also discussed.
\end{abstract}

\setcounter{page}{0}
\thispagestyle{empty}
\pagebreak

\section{Introduction}

Holographic principle provides a powerful tool for calculating the properties of strongly coupled systems\cite{9711200,0903.3246,1106.4324,0909.0518,1304.5908,0904.1975}. According to the principle, a classical weakly coupled gravitational theory is mapped to a strongly coupled large N gauge theory on the boundary. The property of strong-weak duality in this approach makes non-perturbative calculations possible.

Since a system at a critical point can be described by a strongly coupled conformal field theory, the metric in the corresponding classical gravitational theory may be that of Anti-de Sitter space which also possesses scale symmetry. However, in condensed matter physics there are many systems in which space and time scale differently, so it is necessary to introduce the dynamical exponent $z$ in the metric to characterize the Lifshitz scaling\cite{0712.1136,0808.1725,1209.3946,1504.00535,0911.3586,1007.2490,1102.3820,1105.6335,1006.2124,1007.1464,1105.1162,1209.3559,0907.1846,1408.0793}. When the temperature is finite, a black hole metric which is asymptotic to Lifshitz is what we want. Possible models that can yield such solutions, for example, are Einstein-Maxwell-Dilaton model and massive vector field model. Besides, another Maxwell term should be introduced when the system has a finite chemical potential. For more general cases, the metric can also involve a hyperscaling violation factor $\theta$.

Once the background has been set up, one can extract various transport properties such as electric conductivity from it. In the gauge/gravity duality approach, the transport coefficients can be obtained by analyzing the retarded Green's functions from small perturbations about the background. For example, in the case of the AC electric conductivity, one obtain the conductivity by computing the retarded Green's function from the perturbations with time dependence $e^{-i\omega}$, and take $\omega\rightarrow0$ to obtain the DC conductivity.

However, the transport will be divergent if the system is translational invariant. Therefore, some mechanism of momentum relaxation to break translational invariance should be introduced.
 One straightforward approach is to impose inhomogeneous boundary conditions of the bulk fields \cite{1204.0519,1209.1098,1302.6586,1309.4580,1308.0329,1409.6875}.
 Other approaches include introducing extra bulk fields like holographic Q-lattices \cite{1311.3292,1401.5077} and linear massless
  axion \cite{1311.5157,1401.5436,1406.4870,1409.8346,1204.3008,1404.5027,1411.5452,1412.8346,1512.01917,Ge:2016sel,Ge:2016lyn}. The holographic Q-lattice model breaks translational
  invariance by exploiting a continuous global symmetry of bulk gravitational theory while the linear axion model breaks the translational invariance by introducing a spatial dependent
  source of bulk fields. In addition, the momentum dissipation can also be introduced by explicitly breaking the spatial diffeomorphism invariance as the massive gravity
  theory \cite{1301.0537,1306.5792,1308.4970,1310.3832,Hu,Zhang,wu,BP,bkp,matteo,1411.6631,1407.0306,1406.4134}. 

An approach to directly calculate the DC conductivity was introduced in \cite{1406.4742}.
In this approach, the bulk equations of motion can be manipulated into radially independent quantities which are identified with the currents, and the DC conductivity is expressed in terms of the horizon data by analyzing the regularity condition on the horizon. The thermal conductivity and thermoelectric conductivity can also be computed in this approach. In presence of the magnetic field, the DC transport \cite{1502.02631} gives more interesting results such as the Hall angle and the Nernst signal \cite{1406.1659,1502.03789,1502.05386,0704.1160,0706.3215,0706.3228} if the magnetization currents are subtracted carefully to obtain radially independent quantities. A holographic Wilsonian renormalization group approach to momentum dissipated systems were also developed by one of the authors very recently \cite{tgw}.

The subtle points of the transport in Lifshitz spacetime with two gauge fields are that the resulting DC electrical conductivity matrix is hard to interpreted.
 The first gauge field plays the role of an auxiliary field, making the geometry asymptotic Lifshitz, and the second gauge field makes the black hole charged,  playing a role analogous to that of a standard Maxwell field in asymptotically AdS space. The mixture of the two gauge fluctuations leads to a $2\times 2$ conductivity matrix with non-vanishing off-diagonal
components in the absence of external magnetic field, although we consider electric perturbations only along
the $x$-direction.
  In the presence of magnetic fields, the resulting DC electric conductivity becomes a $4\times 4$ matrix.
The physical interpretation of each component is a very tough task, since one does not expect so many components on the dual field theory side. To avoid ambiguities, we shall set the currents induced by the auxiliary gauge field to be vanishing so that
the deduced electrical conductivity matrix is only related to the black hole charges \cite{Ge:2016sel,Ge:2016lyn,lu16}.

In this work, we will construct a dyonic-black-hole-like solution with hyperscaling violating factor and derive various transport coefficients such as electrical and thermoelectric conductivities (see \cite{1602.00556,1602.08476,1601.04732} for related work on transport in Lifshitz geometry and \cite{1603.03029,1708.05691} in hyperscaling violating geometries). This paper is organized as follows. In section 2, we present a dyonic-black-hole-like solution with a dynamic exponent and a hyperscaling violation factor in the Einstein-Maxwell-Dilaton-Axion model. In section 3, we calculate the DC electric conductivity, thermal and thermoelectric conductivities, Hall angle, Nernst signal and Seebeck coefficients in terms of horizon data. In section 4, some special cases are considered. Our conclusion is presented in section 5.

\section{The black hole solutions}

In this section, we will construct a dyonic black hole solution with hyperscaling violating factor in the presence of momentum relaxation. 
In order to obtain the solution which is asymptotic to Lifshitz geometry, we use the Einstein-Maxwell-Dilaton holographic model with two $U(1)$ gauge fields. One gauge field coupled with scalar field is required to generate the Lifshitz scaling, while the other is to provide charge density on the boundary. For further investigations into the transport properties, we also introduce linear axions which leads to momentum dissipation.
Therefore, we consider the following action
\begin{equation}\label{action}
  S=-\frac{1}{16\pi G}\int\md^{4}x\sqrt{-g}\bigg(\mathcal{R}-\frac{1}{2}(\partial\phi)^2
  -\frac{1}{4}\sum_{i=1}^{2}\me^{\lambda_{i}\phi}(F_{i})^{2}
  -\frac{1}{2}\me^{\eta\phi}\sum_{i=1}^{2}(\partial\chi_{i})^{2}
  +\sum_{i=1}^{2}V_{i}\me^{\gamma_{i}\phi}\bigg),
\end{equation}
where $\lambda_{i},\eta,\gamma_{i},V_{i}$ are undetermined constant parameters. We have assumed general gauge-dilaton coupling since this is demanded by the Lifshitz scaling \cite{1105.6335}.  That is to say,  the requirement of having an asymptotically Lifshitz manifold (i.e., for $z\neq 1$) forces a relationship
between the constant of motion associated to $A_i$ and the magnitude of the dilaton field $\phi$.  Otherwise, one would expect to have one free parameter associated to the gauge field, given by the constant of motion , since it appears in the action only through its derivatives. For the relativistic scaling $z=1$, the Maxwell field can be independent of the dilaton field $\phi$.

The equations of motion are given by
\begin{subequations}\label{motionequation}
\begin{gather}
  R\indices{^{\mu}_{\nu}}-\frac{1}{2}\partial^{\mu}\phi\partial_{\nu}\phi -\frac{1}{2}\sum_{i=1}^{2}\me^{\lambda_{i}\phi}(F_{i})^{\mu\rho}(F_{i})_{\nu\rho}
  -\frac{1}{2}\me^{\eta\phi}\sum_{i=1}^{2}\partial^{\mu}\chi_{i}\partial_{\nu}\chi_{i} \nonumber \\
  +\frac{1}{2}\delta^{\mu}_{ \nu}\big(\frac{1}{4}\sum_{i=1}^{2}\me^{\lambda_{i}\phi}(F_{i})^{2}
  +\sum_{i=1}^{2}V_{i}\me^{\gamma_{i}\phi}\big)=0, \label{metricequation}
\end{gather}
\begin{equation}\label{gaugeequation1}
  \partial_{\mu}(\sqrt{-g}\me^{\lambda_{i}\phi}(F_{i})^{\mu\nu})=0,
\end{equation}
\begin{equation}\label{axionequation}
  \partial_{\mu}(\sqrt{-g}\me^{\eta\phi}g^{\mu\nu}\partial_{\nu}\chi_{i})=0,
\end{equation}
\begin{equation}\label{scalarequation1}
   \frac{1}{\sqrt{-g}}\partial_{\mu}(\sqrt{-g}g^{\mu\nu}\partial_{\nu}\phi)
  -\frac{1}{4}\sum_{i=1}^{2}\lambda_{i}\me^{\lambda_{i}\phi}(F_{i})^2
  -\frac{1}{2}\eta\me^{\eta\phi}\sum_{i=1}^{2}(\partial\chi_{i})^2
  +\sum_{i=1}^{2}V_{i}\gamma_{i}\me^{\gamma_{i}\phi}=0.
\end{equation}
\end{subequations}

At the same time, we consider the following ansatz for the metric, gauge fields and axions
\begin{subequations}\label{ansatz}
\begin{equation}\label{metricansatz}
  \md s^{2}=r^{-\theta}\big(-r^{2z}f(r)\,\md t^{2}+\frac{\md r^{2}}{r^{2}f(r)}+r^{2}\,\md x^{2}+r^{2}\,\md y^{2}\big),
\end{equation}
\begin{equation}\label{gaugeansatz}
  A_{1}=a_{1}(r)\,\md t, \qquad
  A_{2}=a_{2}(r)\,\md t+Bx\,\md y,
\end{equation}
\begin{equation}\label{axionsansatz}
  \chi_{1}=kx, \qquad
  \chi_{2}=ky.
\end{equation}
\end{subequations}
where the constants $z$ and $\theta$ are dynamical and hyperscaling violation exponents, respectively. The black hole solution represents effective low temperature geometry,  is not an asymptotically AdS solution and therefore can in principle be interpreted as an IR geometry  embedded in the AdS space. The second gauge field is the physical one which provides the finite chemical potential and the constant $B$ is the magnetic field. Obviously, the ansatz of axions (\ref{axionsansatz}) is just the solution to equations of motion of axions (\ref{axionequation}) if $k$ is just a constant. Besides, the scalar field $\phi$ only depends on the radial coordinate, namely $\phi=\phi(r)$.

Given the ansatz of metric, scalar and gauge fields, the Maxwell equations (\ref{gaugeequation1}) can be recast as
\begin{equation}\label{gaugeequation2}
  \partial_{r}(\sqrt{-g}\me^{\lambda_{i}\phi}(F_{i})^{r\nu})=0,
\end{equation}
or equivalently
\begin{equation}\label{charge}
  q_{i}=J_{i}^{t}=\sqrt{-g}\me^{\lambda_{i}\phi}(F_{i})^{tr}=r^{-z+3}\me^{\lambda_{i}\phi}a_{i}',
\end{equation}
where $q_{i}$ are constants of integration. In  Lifshitz spacetime, $q_{2}$ is interpreted as the charge density in the boundary theory while the ``charge" $q_{1}$ is not really the black hole charge since the first gauge field is only used to support the Lifshitz scaling.

Next, subtracting the $rr$ component from the $tt$ component of Einstein equation (\ref{metricequation})
\begin{equation}\label{tt-rr}
  -\frac{1}{2}(\theta-2)(\theta-2z+2)r^{\theta}f+\frac{1}{2}r^{\theta+2}(\phi')^{2}f=0,
\end{equation}
one can solve the scalar field
\begin{equation}\label{scalarsolution}
  \phi=\sqrt{(\theta-2)(\theta-2z+2)}\ln{r}\equiv\beta\ln{r}.
\end{equation}
The expressions (\ref{charge}) and (\ref{scalarsolution}) lead the $xx$ component Einstein equation (\ref{metricequation}) to
\begin{gather}
  \frac{1}{2}(\theta-2)r^{2\theta-z-1}(r^{-\theta+z+2}f)'-\frac{1}{4}\sum_{i}r^{-\lambda_{i}\beta+2\theta-4}(q_{i})^{2}
  -\frac{1}{4}r^{\lambda_{2}\beta+2\theta-4}B^{2} \nonumber \\
  -\frac{1}{2}r^{\eta\beta+\theta-2}k^{2}+\frac{1}{2}\sum_{i}V_{i}r^{\gamma_{i}\beta}=0, \label{xx}
\end{gather}
from which, by integral, we can solve the function $f$ in terms of some undetermined parameters
\begin{align}\label{fsolution1}
  f= & \sum_{i}\frac{(q_{i})^{2}r^{-\lambda_{i}\beta+\theta-4}}{2(\theta-2)(-\lambda_{i}\beta+z-2)}
  +\frac{B^{2}r^{\lambda_{2}\beta+\theta-4}}{2(\theta-2)(\lambda_{2}\beta+z-2)} \nonumber \\
     & +\frac{k^{2}r^{\eta\beta-2}}{(\theta-2)(\eta\beta-\theta+z)}
     -\sum_{i}\frac{V_{i}r^{\gamma_{i}\beta-\theta}}{(\theta-2)(\gamma_{i}\beta-2\theta+z+2)}
     -mr^{\theta-z-2},
\end{align}
where $m$ is a constant of integration and can be interpreted as the mass of black hole. The condition that metric is the Lifshitz type fixes the parameter $\gamma_{1}$ as
\begin{equation}\label{parametersolution1}
  \gamma_{1}=\frac{\theta}{\beta}.
\end{equation}

The determination of other parameters need the equation of motion of scalar field (\ref{scalarequation1}) in which expressions (\ref{charge}) and (\ref{scalarsolution}) are plugged in
\begin{gather}
  \beta r^{2\theta-z-1}(r^{-\theta+z+2}f)'+\frac{1}{2}\sum_{i}\lambda_{i} r^{-\lambda_{i}\beta+2\theta-4}(q_{i})^{2}
  -\frac{1}{2}\lambda_{2}r^{\lambda_{2}\beta+2\theta-4}B^{2} \nonumber\\
  -\eta r^{\eta\beta+\theta-2}k^{2}+\sum_{i}V_{i}\gamma_{i}r^{\gamma_{i}\beta}=0. \label{scalarequation2}
\end{gather}
Combining (\ref{xx}) with (\ref{scalarequation2}) and eliminating the function $f$, one obtains
\begin{gather}\label{parameterequation}
  \frac{1}{2}\sum_{i}[\beta+\lambda_{i}(\theta-2)]r^{-\lambda_{i}\beta+2\theta-4}(q_{i})^{2}
  +\frac{1}{2}[\beta-\lambda_{2}(\theta-2)]r^{\lambda_{2}\beta+2\theta-4}B^{2} \nonumber\\
  +[\beta-\eta(\theta-2)]r^{\eta\beta+\theta-2}k^{2}
  +\sum_{i}V_{i}[\gamma_{i}(\theta-2)-\beta]r^{\gamma_{i}\beta}=0.
\end{gather}
 \textbf{First class of the solution--} Since $q_{2}$ and $k$ can be arbitrary value, their coefficients should be zero and we can obtain the values of $\lambda_{2}$ and $\eta$. Although $B$ is also arbitrary, its coefficient has no more parameter after $\lambda_{2}$ is determined, so we let the exponentials of the terms which contain $B$ and $V_{2}$ to be equal and let their coefficients be canceled with each other. Meanwhile use the same way to deal with the remaining terms so that the equation is satisfied. The results are
\begin{gather}
  \lambda_{2}=\frac{\beta}{2-\theta}, \qquad
  \eta=\frac{\beta}{\theta-2}, \qquad
  \lambda_{1}=\frac{\theta-4}{\beta}, \qquad
  \gamma_{2}=\frac{\theta+2z-6}{\beta}, \nonumber \\
  (q_{1})^{2}=\frac{2V_{1}(z-1)}{z-\theta+1}, \qquad
  V_{2}=\frac{B^{2}(2z-\theta-2)}{4(z-2)}. \label{parametersolution2}
\end{gather}
Then we can plug all parameters (\ref{parametersolution1}) (\ref{parametersolution2}) into the expression (\ref{fsolution1}) and obtain the final result of function
\begin{equation}\label{fsolution2}
  f=1+\frac{(q_{2})^{2}r^{2\theta-2z-2}}{2(\theta-2)(\theta-z)}+\frac{B^{2}r^{2z-6}}{4(z-2)(3z-\theta-4)}
  -\frac{k^{2}r^{\theta-2z}}{(\theta-2)(z-2)}-mr^{\theta-z-2}.
\end{equation}
The constant term is set to be one, as long as we demand
\begin{equation}\label{v1solution}
  V_{1}=(z-\theta+1)(z-\theta+2).
\end{equation}
Also, we can obtain the solution of gauge fields using (\ref{charge}), (\ref{scalarsolution}) and(\ref{parametersolution2})
\begin{equation}\label{gaugesolution}
  a_{1}=\frac{q_{1}r^{z-\theta+2}}{z-\theta+2}+\mu_{1}, \qquad
  a_{2}=\frac{q_{2}r^{\theta-z}}{\theta-z}+\mu_{2},
\end{equation}
where $\mu_{i}$ are constants of integration and the physical one $\mu_{2}$ is the chemical potential. In order that the chemical potential meaning makes sense, the dynamical and hyperscaling violation exponents should satisfy $\theta<z$. So far we can see that the introducing one more gauge field coupled to the scalar field indeed is a way to solve the anisotropic scaling, since we can check that the $tt$ component of Einstein equation (\ref{metricequation}) is automatically satisfied when all above constraints are imposed.

Using the definition of horizon $f(r_{+})=0$ , we can express the mass-parameter $m$ in terms of $r_{+}$
\begin{equation}\label{mass}
  m=r_{+}^{z-\theta+2}+\frac{(q_{2})^{2}r_{+}^{\theta-z}}{2(\theta-2)(\theta-z)}
  +\frac{B^{2}r_{+}^{3z-\theta-4}}{4(z-2)(3z-\theta-4)}-\frac{k^{2}r_{+}^{-z+2}}{(\theta-2)(z-2)},
\end{equation}
and further calculate the Hawking temperature
\begin{equation}\label{temperature}
  T=\frac{1}{4\pi}\bigg[(z-\theta+2)r_{+}^{z}+\frac{(q_{2})^{2}r_{+}^{2\theta-z-2}}{2(\theta-2)}
  +\frac{B^{2}r_{+}^{3z-6}}{4(z-2)}+\frac{k^{2}r_{+}^{\theta-z}}{\theta-2}\bigg].
\end{equation}

 \textbf{Second class of the solution--} In order to obtain (\ref{parametersolution2}), we have supposed that the requirement of charge $q_{2}$ being free is fulfilled first and then used the term containing $V_{2}$ to offset the term containing the free $B$. In fact, we can reverse these two steps, namely let the coefficient of $B$ be zero first and then demand that the exponentials of the terms which contain $q_{2}$ and $V_{2}$ to be equal and their coefficients to be canceled with each other. The results will be slightly changed
\begin{gather}
  \lambda_{2}^{*}=\frac{\beta}{\theta-2}, \qquad
  \gamma_{2}^{*}=\frac{\theta+2z-6}{\beta}, \qquad
  V_{2}^{*}=\frac{(q_{2})^{2}(2z-\theta-2)}{4(z-2)}. \label{parametersolution3}
\end{gather}
Other parameters will be the same since they are determined in the same way. The corresponding (\ref{fsolution2}) becomes
\begin{equation}\label{fsolution3}
  f^{*}=1+\frac{(q_{2})^{2}r^{2z-6}}{4(z-2)(3z-\theta-4)}+\frac{B^{2}r^{2\theta-2z-2}}{2(\theta-2)(\theta-z)}
  -\frac{k^{2}r^{\theta-2z}}{(\theta-2)(z-2)}-mr^{\theta-z-2}.
\end{equation}
The difference between (\ref{fsolution2}) and (\ref{fsolution3}) is the interchange of the coefficients and the exponentials of the terms containing $q_{2}$ and $B$. Then the mass and the Hawking temperature are
\begin{equation}\label{mass2}
  m^{*}=r_{+}^{z-\theta+2}+\frac{(q_{2})^{2}r_{+}^{3z-\theta-4}}{4(z-2)(3z-\theta-4)}
  +\frac{B^{2}r_{+}^{\theta-z}}{2(\theta-2)(\theta-z)}-\frac{k^{2}r_{+}^{-z+2}}{(\theta-2)(z-2)},
\end{equation}
\begin{equation}\label{temperature2}
  T^{*}=\frac{1}{4\pi}[(z-\theta+2)r_{+}^{z}+\frac{(q_{2})^{2}r_{+}^{3z-6}}{4(z-2)}
  +\frac{B^{2}r_{+}^{2\theta-z-2}}{2(\theta-2)}+\frac{k^{2}r_{+}^{\theta-z}}{\theta-2}].
\end{equation}

\section{Thermo-electric transport}

Now  we begin to calculate the electric conductivity $\sigma$ and thermoelectric conductivity $\alpha$ in terms of horizon data. For notation simplicity, we rewrite the action (\ref{action}) and the ansatz of the metric (\ref{metricansatz}) as
\begin{equation}\label{Action}
    S=\int\md^{4}x\sqrt{-g}\big(\mathcal{R}-\frac{1}{2}(\partial\phi)^2
  -\frac{1}{4}\sum_{i=1}^{2}Z_{i}(F_{i})^{2}
  -\frac{1}{2}\Phi\sum_{i=1}^{2}(\partial\chi_{i})^{2}
  +\sum_{i=1}^{2}V_{i}\big),
\end{equation}
\begin{equation}\label{Metricansatz}
  \md s^{2}=-U(r)\,\md t^{2}+V(r)\,\md r^{2}+W(r)\,\md x^{2}+W(r)\,\md y^{2},
\end{equation}
and maintain the remaining ansatz (\ref{gaugeansatz}) and (\ref{axionsansatz}).

In order to compute conductivities, we consider the following small perturbations around the background solutions, just as \cite{1406.4742}
\begin{gather}\label{perturbation}
  \delta g_{tx}=h_{tx}(r), \qquad \delta g_{ty}=h_{ty}(r), \qquad
  \delta g_{rx}=h_{rx}(r), \qquad \delta g_{ry}=h_{ry}(r), \nonumber \\
  \delta A_{ix}=b_{ix}(r)-E_{ix}t, \qquad \delta A_{iy}=b_{iy}(r)-E_{iy}t, \nonumber \\
  \delta \chi_{1}=\varphi_{x}(r), \qquad \delta \chi_{2}=\varphi_{y}(r),
\end{gather}
where $E_{ix}$ and $E_{iy}$ are constants, $i$ takes $1$ or $2$. For a complete understanding of the holographic interpretation, we would like to address the UV asymptotic  of electric-magnetic perturbations. The larger-$r$ asymptotic behavior for $\delta A_{ix}$ for the first class of the solution are given by
\bea
\delta A_{1x}=c_{10}\bigg[1+\frac{c_{11}(\omega)}{r^{z-4+\theta}}+...\bigg],\\
\delta A_{2x}=c_{20}\bigg[1+\frac{c_{22}(\omega)}{r^{3z-2+\theta}}+...\bigg].
\eea
The same is for  $\delta A_{iy}$.
Note that  in order for $c_{10}$ to be regular asymptotically, one must have $c_{11}  = 0$ when
$z-4+\theta < 0$.  In the $\omega\rightarrow 0$ limit, we can define
\be
\tilde{\sigma}_{ij}=-\lim_{\omega\rightarrow 0}\frac{c_{ij}(\omega)}{i\omega}, ~~~i=1,2.
\ee
One can prove that $\tilde{\sigma}_{ij}$ is exactly the DC conductivity matrix as given in (39). The similar discussions are given in  \cite{lu16,Ge:2016lyn}.

Then we can obtain the linearized Maxwell equation in the form of
\begin{equation}\label{linearizedgauge}
  \partial_{r}(\sqrt{-g}Z_{i}(F_{i})^{nr})=0.
\end{equation}
According to the holographic principle, the electric current density $J^{n}=(J^{t},J^{x},J^{y})$ has the form
\begin{equation}\label{eletriccurrent1}
  J^{n}=\sqrt{-g}ZF^{nr},
\end{equation}
which is calculated at the boundary $r\rightarrow\infty$. Comparing these two expressions , we can conclude that $J^{n}$ is a conserved quantity along radial direction and can be evaluated at arbitrary value of $r$.

It is easier to do the computation at the horizon. Therefore we make the coordinate transformation $v=t+\int\md r\sqrt{\frac{V}{U}}$ so that the background metric is explicitly regular at the horizon, namely $\md s^{2}=-U\md v^{2}-2\sqrt{UV}\md v\md r+W\md x^{2}+W\md y^{2}$. Now the perturbed metric has additional terms
\begin{equation}\label{transformedmertic}
  h_{tx}\md v\md x+h_{ty}\md v\md y
  +(h_{rx}-\sqrt{\frac{V}{U}}g_{hx})\md r\md x+(h_{ry}-\sqrt{\frac{V}{U}}g_{hy})\md r\md y.
\end{equation}
In order to ensure the regularity of the perturbed metric at the horizon, we demand the following relation at the horizon
\begin{equation}\label{perturbationrelation1}
  h_{rx}\sim\sqrt{\frac{V}{U}}h_{tx}, \qquad
  h_{ry}\sim\sqrt{\frac{V}{U}}h_{ty}.
\end{equation}
The gauge fields should also be regular at the horizon, so from the expression of the perturbed gauge fields
\begin{equation}\label{transformationgauge}
  A_{ix}=b_{ix}-E_{ix}v+E_{ix}\int\md r\sqrt{\frac{V}{U}}, \qquad
  A_{iy}=b_{iy}-E_{iy}v+E_{iy}\int\md r\sqrt{\frac{V}{U}},
\end{equation}
following relation should be imposed at the horizon as well
\begin{equation}\label{perturbationrelation2}
  b_{ix}'\sim-\sqrt{\frac{V}{U}}E_{ix}, \qquad b_{iy}'\sim-\sqrt{\frac{V}{U}}E_{iy}.
\end{equation}

To fix the perturbations, we need the linearized Einstein equations
\begin{subequations}\label{linearizedmetric}
\begin{gather}
  \frac{1}{2UV}h_{tx}''-\frac{1}{4UV}(\frac{U'}{U}+\frac{V'}{V})h_{tx}'
  +\frac{1}{4UVW}[W'(\frac{U'}{U}+\frac{V'}{V})-2W'']h_{tx}-\frac{Z_{2}B^{2}}{2UW^{2}}h_{tx} \nonumber \\
  +\frac{Z_{2}Ba_{2}'}{2UVW}h_{ry}+\frac{Z_{2}a_{2}'}{2UV}b_{2x}'
  -\frac{Z_{2}B}{2UW}E_{2y}+\frac{Z_{1}a_{1}'}{2UV}b_{1x}'-\frac{\Phi k^{2}}{2UW}h_{tx}=0, \label{tx}
\end{gather}
\begin{gather}
  \frac{1}{2UV}h_{ty}''-\frac{1}{4UV}(\frac{U'}{U}+\frac{V'}{V})h_{ty}'
  +\frac{1}{4UVW}[W'(\frac{U'}{U}+\frac{V'}{V})-2W'']h_{ty}-\frac{Z_{2}B^{2}}{2UW^{2}}h_{ty} \nonumber \\
  -\frac{Z_{2}Ba_{2}'}{2UVW}h_{rx}+\frac{Z_{2}a_{2}'}{2UV}b_{2y}'
  +\frac{Z_{2}B}{2UW}E_{2x}+\frac{Z_{1}a_{1}'}{2UV}b_{1y}'-\frac{\Phi k^{2}}{2UW}h_{ty}=0, \label{ty}
\end{gather}
\begin{equation}\label{rx}
  \frac{Z_{2}B^{2}}{2VW^{2}}h_{rx}-\frac{Z_{2}Ba_{2}'}{2UVW}h_{ty}-\frac{Z_{2}B}{2VW}b_{2y}'
  +\frac{Z_{2}a_{2}'}{2UV}E_{2x}+\frac{Z_{1}a_{1}'}{2UV}E_{1x}+\frac{\Phi k^{2}}{2VW}h_{rx}
  -\frac{\Phi k}{2V}\varphi_{x}'=0,
\end{equation}
\begin{equation}\label{ry}
  \frac{Z_{2}B^{2}}{2VW^{2}}h_{ry}+\frac{Z_{2}Ba_{2}'}{2UVW}h_{tx}+\frac{Z_{2}B}{2VW}b_{2x}'
  +\frac{Z_{2}a_{2}'}{2UV}E_{2y}+\frac{Z_{1}a_{1}'}{2UV}E_{1y}+\frac{\Phi k^{2}}{2VW}h_{ry}
  -\frac{\Phi k}{2V}\varphi_{y}'=0.
\end{equation}
\end{subequations}
Taking into account the horizon values (\ref{charge}) (\ref{perturbationrelation1}) (\ref{perturbationrelation2}), either $tx$ (\ref{tx}) and $ty$ (\ref{ty}) or $rx$  (\ref{rx}) and $ry$ (\ref{ry}) components of Einstein equation will give
\begin{subequations}\label{pertubationequation}
\begin{equation}
  (Z_{2}B^{2}+W\Phi k^{2})h_{tx}-q_{2}Bh_{ty}=-Wq_{2}E_{2x}-Wq_{1}E_{1x}-WZ_{2}BE_{2y},
\end{equation}
\begin{equation}
  q_{2}Bh_{tx}+(Z_{2}B^{2}+W\Phi k^{2})h_{ty}=-Wq_{2}E_{2y}-Wq_{1}E_{1y}+WZ_{2}BE_{2x}.
\end{equation}
\end{subequations}
The solutions of the linearized Einstein equations are
\begin{subequations}
\begin{align}\label{htx}
  h_{tx}= & -[W^{2}\Phi k^{2}q_{2}E_{2x}+WB(q_{2}^{2}+Z_{2}^{2}B^{2}+Z_{2}W\Phi k^{2})E_{2y} \nonumber \\
          & +Wq_{1}(Z_{2}B^{2}+W\Phi k^{2})E_{1x}+Wq_{1}q_{2}BE_{1y}]/[(Z_{2}B^{2}+W\Phi k^{2})^{2}+q_{2}^{2}B^{2}],
\end{align}
\begin{align}\label{hty}
  h_{ty}= & [WB(q_{2}^{2}+Z_{2}^{2}B^{2}+Z_{2}W\Phi k^{2})E_{2x}-W^{2}\Phi k^{2}q_{2}E_{2y} \nonumber \\
          & +Wq_{1}q_{2}BE_{1x}-Wq_{1}(Z_{2}B^{2}+W\Phi k^{2})E_{1y}]/[(Z_{2}B^{2}+W\Phi k^{2})^{2}+q_{2}^{2}B^{2}].
\end{align}
\end{subequations}

From the linearized Maxwell equation (\ref{linearizedgauge}), we can obtain four radially independent quantities
\begin{subequations}
\begin{align}\label{J1x}
  J_{1x}= & \left.Z_{1}E_{1x}-\frac{q_{1}}{W}h_{tx}\right|_{r=r_{+}} \nonumber \\
        = & Z_{1}E_{1x}+q_{1}[W\Phi k^{2}q_{2}E_{2x}+B(q_{2}^{2}+Z_{2}^{2}B^{2}+Z_{2}W\Phi k^{2})E_{2y} \nonumber \\
          & +q_{1}(Z_{2}B^{2}+W\Phi k^{2})E_{1x}+q_{1}q_{2}BE_{1y}]/[(Z_{2}B^{2}+W\Phi k^{2})^{2}+q_{2}^{2}B^{2}],
\end{align}
\begin{align}\label{J1y}
  J_{1y}= & \left.Z_{1}E_{1y}-\frac{q_{1}}{W}h_{ty}\right|_{r=r_{+}} \nonumber \\
        = & Z_{1}E_{1y}-q_{1}[B(q_{2}^{2}+Z_{2}^{2}B^{2}+Z_{2}W\Phi k^{2})E_{2x}-W\Phi k^{2}q_{2}E_{2y} \nonumber \\
          & +q_{1}q_{2}BE_{1x}-q_{1}(Z_{2}B^{2}+W\Phi k^{2})E_{1y}]/[(Z_{2}B^{2}+W\Phi k^{2})^{2}+q_{2}^{2}B^{2}].
\end{align}
\begin{align}\label{J2x}
  J_{2x}= & \left.Z_{2}E_{2x}-\frac{q_{2}}{W}h_{tx}-\frac{Z_{2}B}{W}h_{ty}\right|_{r=r_{+}} \nonumber \\
        = & [W\Phi k^{2}(q_{2}^{2}+Z_{2}^{2}B^{2}+WZ_{2}\Phi k^{2})E_{2x}+q_{2}B(q_{2}^{2}+Z_{2}^{2}B^{2}
          +2WZ_{2}\Phi k^{2})E_{2y} \nonumber \\
          & +W\Phi k^{2}q_{2}q_{1}E_{1x}+q_{1}B(q_{2}^{2}+Z_{2}^{2}B^{2}
          +WZ_{2}\Phi k^{2})E_{1y}]/[q_{2}^{2}B_{2}+(Z_{2}B^{2}+W\Phi k^{2})^2]
\end{align}
\begin{align}\label{J2y}
  J_{2y}= & \left.Z_{2}E_{2y}-\frac{q_{2}}{W}h_{ty}+\frac{Z_{2}B}{W}h_{tx}\right|_{r=r_{+}} \nonumber \\
        = & [-q_{2}B(q_{2}^{2}+Z_{2}^{2}B^{2}+2WZ_{2}\Phi k^{2})E_{2x}
          +W\Phi k^{2}(q_{2}^{2}+Z_{2}^{2}B^{2}+WZ_{2}\Phi k^{2})E_{2y} \nonumber \\
          & -q_{1}B(q_{2}^{2}+Z_{2}^{2}B^{2}+WZ_{2}\Phi k^{2})E_{1x}
          +W\Phi k^{2}q_{2}q_{1}E_{1y}]/[q_{2}^{2}B_{2}+(Z_{2}B^{2}+W\Phi k^{2})^2]
\end{align}
\end{subequations}
If we just take the derivative $\sigma_{iajb}=\frac{\partial J_{ia}}{\partial E_{jb}}$, we will obtain following 16-quantities
\begin{flalign*}
  \sigma_{1x1x} & =Z_{1}+\frac{q_{1}^{2}(Z_{2}B^{2}+W\Phi k^{2})}{(Z_{2}B^{2}+W\Phi k^{2})^{2}+q_{2}^{2}B^{2}},
  & \sigma_{1x1y} & =\frac{q_{1}^{2}q_{2}B}{(Z_{2}B^{2}+W\Phi k^{2})^{2}+q_{2}^{2}B^{2}}, \\
  \sigma_{1x2x} & =\frac{q_{1}W\Phi k^{2}q_{2}}{(Z_{2}B^{2}+W\Phi k^{2})^{2}+q_{2}^{2}B^{2}},
  & \sigma_{1x2y} & =\frac{q_{1}B(q_{2}^{2}+Z_{2}^{2}B^{2}+Z_{2}W\Phi k^{2})}{(Z_{2}B^{2}+W\Phi k^{2})^{2}+q_{2}^{2}B^{2}}, \\
  \sigma_{1y1x} & =\frac{-q_{1}^{2}q_{2}B}{(Z_{2}B^{2}+W\Phi k^{2})^{2}+q_{2}^{2}B^{2}},
  & \sigma_{1y1y} & =Z_{1}+\frac{q_{1}^{2}(Z_{2}B^{2}+W\Phi k^{2})}{(Z_{2}B^{2}+W\Phi k^{2})^{2}+q_{2}^{2}B^{2}}, \\
  \sigma_{1y2x} & =-\frac{q_{1}B(q_{2}^{2}+Z_{2}^{2}B^{2}+Z_{2}W\Phi k^{2})}{(Z_{2}B^{2}+W\Phi k^{2})^{2}+q_{2}^{2}B^{2}},
  & \sigma_{1y2y} & =\frac{q_{1}W\Phi k^{2}q_{2}}{(Z_{2}B^{2}+W\Phi k^{2})^{2}+q_{2}^{2}B^{2}}, \\
  \sigma_{2x1x} & =\frac{W\Phi k^{2}q_{2}q_{1}}{(Z_{2}B^{2}+W\Phi k^{2})^{2}+q_{2}^{2}B^{2}},
  & \sigma_{2x1y} & =\frac{q_{1}B(q_{2}^{2}+Z_{2}^{2}B^{2}+WZ_{2}\Phi k^{2})}{(Z_{2}B^{2}+W\Phi k^{2})^{2}+q_{2}^{2}B^{2}}, \\
  \sigma_{2x2x} & =\frac{W\Phi k^{2}(q_{2}^{2}+Z_{2}^{2}B^{2}+WZ_{2}\Phi k^{2})}{(Z_{2}B^{2}+W\Phi k^{2})^{2}+q_{2}^{2}B^{2}},
  & \sigma_{2x2y} & =\frac{q_{2}B(q_{2}^{2}+Z_{2}^{2}B^{2}+2WZ_{2}\Phi k^{2})}{(Z_{2}B^{2}+W\Phi k^{2})^{2}+q_{2}^{2}B^{2}}, \\
  \sigma_{2y1x} & =-\frac{q_{1}B(q_{2}^{2}+Z_{2}^{2}B^{2}+WZ_{2}\Phi k^{2})}{(Z_{2}B^{2}+W\Phi k^{2})^{2}+q_{2}^{2}B^{2}},
  & \sigma_{2y1y} & =\frac{W\Phi k^{2}q_{2}q_{1}}{(Z_{2}B^{2}+W\Phi k^{2})^{2}+q_{2}^{2}B^{2}}, \\
  \sigma_{2y2x} & =-\frac{q_{2}B(q_{2}^{2}+Z_{2}^{2}B^{2}+2WZ_{2}\Phi k^{2})}{(Z_{2}B^{2}+W\Phi k^{2})^{2}+q_{2}^{2}B^{2}},
  & \sigma_{2y2y} & =\frac{W\Phi k^{2}(q_{2}^{2}+Z_{2}^{2}B^{2}+WZ_{2}\Phi k^{2})}{(Z_{2}B^{2}+W\Phi k^{2})^{2}+q_{2}^{2}B^{2}}.
\end{flalign*}

However, as $A_{2}$ is a physical gauge field while $A_{1}$ is only an auxiliary field, the only physical electrical fields are $E_{2x}$ and $E_{2y}$, and the physical electrical currents are $J_{2x},J_{2y}$. Therefore, we can identify $\sigma_{2x2x},\sigma_{2x2y},\sigma_{2y2x},\sigma_{2y2y}$ as conductivities, but have no idea about the rest. These four conductivities are the same as those in \cite{1502.03789} which is based on asymptotic AdS background. We can infer further that the thermoelectric conductivities and thermal conductivities are the same as the old results as long as we do not include the contributions from $A_{1}$.

Or we can make some efforts to eliminate the explicit dependence on $A_{1}$. Let $J_{1x}=0$, $J_{1y}=0$ and solve for $E_{1x},E_{1y}$ so that we can obtain the expressions of electric currents into which $E_{1x},E_{1y}$ are substituted for. The results also allow us to compute the electrical conductivity
\begin{subequations}\label{electricalconductivity}
\begin{align}
  \sigma_{xx}=\frac{\partial J_{2x}}{\partial E_{2x}}
             =\frac{(q_{1}^{2}+Z_{1}W\Phi k^{2})[Z_{1}(q_{2}^{2}+Z_{2}^{2}B^{2})+Z_{2}(q_{1}^{2}+Z_{1}W\Phi k^{2})]}
             {[q_{1}^{2}+Z_{1}(Z_{2}B^{2}+W \Phi k^{2})]^{2}+Z_{1}^{2}q_{2}^{2}B^{2}},
\end{align}
\begin{align}
  \sigma_{xy}=\frac{\partial J_{2x}}{\partial E_{2y}}
             =\frac{Z_{1}q_{2}B[Z_{1}(q_{2}^{2}+Z_{2}^{2}B^{2})+2Z_{2}(q_{1}^{2}+Z_{1}W\Phi k^{2})]}
             {[q_{1}^{2}+Z_{1}(Z_{2}B^{2}+W \Phi k^{2})]^{2}+Z_{1}^{2}q_{2}^{2}B^{2}},
\end{align}
\begin{align}
  \sigma_{yx}=\frac{\partial J_{2y}}{\partial E_{2x}}
             =-\frac{Z_{1}q_{2}B[Z_{1}(q_{2}^{2}+Z_{2}^{2}B^{2})+2Z_{2}(q_{1}^{2}+Z_{1}W\Phi k^{2})]}
             {[q_{1}^{2}+Z_{1}(Z_{2}B^{2}+W \Phi k^{2})]^{2}+Z_{1}^{2}q_{2}^{2}B^{2}},
\end{align}
\begin{align}
  \sigma_{yy}=\frac{\partial J_{2y}}{\partial E_{2y}}
             =\frac{(q_{1}^{2}+Z_{1}W\Phi k^{2})[Z_{1}(q_{2}^{2}+Z_{2}^{2}B^{2})+Z_{2}(q_{1}^{2}+Z_{1}W\Phi k^{2})]}
             {[q_{1}^{2}+Z_{1}(Z_{2}B^{2}+W \Phi k^{2})]^{2}+Z_{1}^{2}q_{2}^{2}B^{2}}.
\end{align}
\end{subequations}
We can check that we still have $\sigma_{yx}=-\sigma_{xy}$ and $\sigma_{yy}=\sigma_{xx}$.

In this situation, we can express the Hall angle as
\begin{equation}
  \theta_{H}\equiv\frac{\sigma_{xy}}{\sigma_{xx}}=\frac{Z_{1}q_{2}B[Z_{1}(q_{2}^{2}+Z_{2}^{2}B^{2})+2Z_{2}(q_{1}^{2}+Z_{1}W\Phi k^{2})]}
  {(q_{1}^{2}+Z_{1}W\Phi k^{2})[Z_{1}(q_{2}^{2}+Z_{2}^{2}B^{2})+Z_{2}(q_{1}^{2}+Z_{1}W\Phi k^{2})]}.
\end{equation}
Another interesting quantity is the magnetoresistance, which is given as
\begin{equation}
  \frac{\Delta\rho}{\rho}=\frac{\rho_{xx}(B)-\rho_{xx}(0)}{\rho_{xx}(0)}
  =\frac{Z_{1}Z_{2}^{3}B^{2}(q_{1}^{2}+Z_{1}W\Phi k^{2})}{(Z_{1}q_{2}^{2}+Z_{2}q_{1}^{2}+Z_{1}Z_{2}W\Phi k^{2})^{2}+Z_{1}^{2}Z_{2}^{2}q_{2}^{2}B^{2}}
\end{equation}
In order to obtain the thermoelectrical conductivities, we need the expressions for heat currents which should be conserved quantities of Einstein equations. The situation is analogous to the electric current. Here we use the two-form in \cite{1406.4742}, so the heat currents are
\begin{subequations}\label{heatcurrent}
\begin{align}
  Q_{x}=\left.\frac{U^{2}}{\sqrt{UV}}(\frac{htx}{U})'-\sum_{i=1}^{2}a_{i}J_{ix}\right|_{r=r_{+}}
       =\left.-\frac{U'h_{tx}}{\sqrt{UV}}\right|_{r=r_{+}},
\end{align}
\begin{align}
  Q_{y}=\left.\frac{U^{2}}{\sqrt{UV}}(\frac{hty}{U})'-\sum_{i=1}^{2}a_{i}J_{iy}\right|_{r=r_{+}}
       =\left.-\frac{U'h_{ty}}{\sqrt{UV}}\right|_{r=r_{+}}.
\end{align}
\end{subequations}
Then the thermoelectrical conductivities are
\begin{subequations}\label{thermoelectricalconductivity}
\begin{align}
  \alpha_{xx}=\frac{1}{T}\frac{\partial Q_{x}}{\partial E_{2x}}
             =\frac{4\pi Z_{1}Wq_{2}(q_{1}^{2}+Z_{1}W\Phi k^{2})}
             {[q_{1}^{2}+Z_{1}(Z_{2}B^{2}+W \Phi k^{2})]^{2}+Z_{1}^{2}q_{2}^{2}B^{2}},
\end{align}
\begin{align}
  \alpha_{xy}=\frac{1}{T}\frac{\partial Q_{x}}{\partial E_{2y}}
             =\frac{4\pi Z_{1}WB[Z_{1}(q_{2}^{2}+Z_{2}^{2}B^{2})+Z_{2}(q_{1}^{2}+Z_{1}W\Phi k^{2})]}
             {[q_{1}^{2}+Z_{1}(Z_{2}B^{2}+W \Phi k^{2})]^{2}+Z_{1}^{2}q_{2}^{2}B^{2}},
\end{align}
\begin{align}
  \alpha_{yx}=\frac{1}{T}\frac{\partial Q_{y}}{\partial E_{2x}}
             =-\frac{4\pi Z_{1}WB[Z_{1}(q_{2}^{2}+Z_{2}^{2}B^{2})+Z_{2}(q_{1}^{2}+Z_{1}W\Phi k^{2})]}
             {[q_{1}^{2}+Z_{1}(Z_{2}B^{2}+W \Phi k^{2})]^{2}+Z_{1}^{2}q_{2}^{2}B^{2}},
\end{align}
\begin{align}
  \alpha_{yy}=\frac{1}{T}\frac{\partial Q_{y}}{\partial E_{2y}}
             =\frac{4\pi Z_{1}Wq_{2}(q_{1}^{2}+Z_{1}W\Phi k^{2})}
             {[q_{1}^{2}+Z_{1}(Z_{2}B^{2}+W \Phi k^{2})]^{2}+Z_{1}^{2}q_{2}^{2}B^{2}}.
\end{align}
\end{subequations}

We can obtain other interesting transport coefficients associated both the electric and heat currents,
the Seebeck coefficient
\begin{equation}
S\equiv\frac{\alpha_{xx}}{\sigma_{xx}}=\frac{4\pi Z_{1}Wq_{2}}{Z_{1}(q^{2}_{2}+Z^{2}_{2}B^{2})+Z_{2}(q^{2}_{1}+Z_{1}W\Phi k^{2})}
\end{equation}
The Nernst signal is then ready to be calculated
\begin{equation}
e_{N}\equiv (\sigma^{-1}\cdot \alpha)_{xy}=\frac{4\pi Z_{1}Z_{2}^{2}WB(q_{1}^{2}+Z_{1}W\Phi k^{2})}{(Z_{1}q_{2}^{2}+Z_{2}q_{1}^{2}+Z_{1}Z_{2}W\Phi k^{2})^{2}+Z_{1}^{2}Z_{2}^{2}q_{2}^{2}B^{2}}.
\end{equation}
The backreacted DBI magnetotransport with momentum dissipation was discussed in \cite{1707.01505}.
\section{Special cases}

So far we have derived the general expressions for conductivities. Now we analysis some special cases. Since we have two general background metric, the discussion will be divided into two parts first. Then we will consider another special case without momentum dissipation.

\textbf{$\bullet$ First class solution with momentum dissipation--}   Now we take (\ref{fsolution2}) as the background metric, then taking (\ref{parametersolution2}) into account, then we have
\begin{equation}\label{actionprameter1}
  (q_{1})^{2}=2(z-1)(z-\theta+2), \quad
  \Phi=r^{\theta-2z+2}, \quad
  Z_{1}=r^{\theta-4}, \quad
  Z_{2}=r^{2z-\theta-2}, \quad
  W=r^{-\theta+2}.
\end{equation}
The simplest case will be $z=1,\theta=0$, in which the metric will return to asymptotic AdS with one charge
\begin{equation}
  f=1+\frac{(q_{2})^{2}}{4r^{4}}+\frac{B^{2}}{4r^{4}}-\frac{k^{2}}{2r^{2}}-\frac{m}{r^{3}}.
\end{equation}
Direct calculation leads to the temperature
\begin{equation}
  T=\frac{1}{4\pi}\bigg(3r_{+}-\frac{(q_{2})^{2}}{4r_{+}^{3}}-\frac{B^{2}}{4r_{+}^{3}}-\frac{k^{2}}{2r_{+}}\bigg).
\end{equation}
We emphasize that once we switch the asymptotic structure from an AdS to a Lifshitz one and turn on the perturbation $\delta A_{(1)x}$, it could not have a continuous limit back to the perturbation considered in the Reissner-Nordstr$\rm\ddot{o}$m-AdS spacetime by simply taking  $z\rightarrow 1$, $\theta \rightarrow 0$ and $q_1 \rightarrow 0$ limit.
From (\ref{actionprameter1}) we can see that taking $z\rightarrow 1$, $\theta\rightarrow 0$ leads to $q_1\rightarrow 0$, but the quantity $\sigma_{1x1x}=\sigma_{1y1y}=r^{-4}_{+}$ is not vanishing. However, if we set $z=1$ and $\theta=0$ from the very beginning in the action (\ref{Action}),
the auxiliary gauge field $F_{(1)rt}$  naturally does not appear and the black hole solution is the Reissner-Nordstr$\rm\ddot{o}$m-AdS (RN-AdS) metric. So we have
a discontinuity in the $z\rightarrow 1$,  $\theta\rightarrow 0$  and $q_1 \rightarrow 0$ limit.

For RN-AdS black hole with linear axions,  the electric and thermoelectric conductivities are given by
\begin{gather}\
  \sigma_{xx}=\frac{r_{+}^{2}k^{2}(q_{2}^{2}+B^{2}+r_{+}^{2}k^{2})}{(B^{2}+r_{+}^{2}k^{2})^{2}+q_{2}^{2}B^{2}}, \nonumber \\
  \sigma_{xy}=\frac{q_{2}B(q_{2}^{2}+B^{2}+2r_{+}^{2}k^{2})}{(B^{2}+r_{+}^{2}k^{2})^{2}+q_{2}^{2}B^{2}}, \nonumber \\
  \alpha_{xx}=\frac{4\pi r_{+}^{4}q_{2}k^{2}}{(B^{2}+r_{+}^{2}k^{2})^{2}+q_{2}^{2}B^{2}}, \nonumber \\
  \alpha_{xy}=\frac{4\pi r_{+}^{2}B(q_{2}^{2}+B^{2}+r_{+}^{2}k^{2})}{(B^{2}+r_{+}^{2}k^{2})^{2}+q_{2}^{2}B^{2}}.
\end{gather}

Of course, we are more interested  in  non-vanishing hyperscaling factor cases, for example $z=1$ and $\theta=1$.  One may notice that the charge term in the metric will be divergent if one plug in the values of the exponents directly (i.e. $z=1$ and $\theta=1$). In order to fix the problem, we rewrite the charge term as
\begin{equation}
  \frac{(q_{2})^{2}r^{2\theta-2z-2}}{2(\theta-2)(\theta-z)}=\frac{(q_{2})^{2}r^{\theta-z}r^{\theta-z-2}}{2(\theta-2)(\theta-z)},
\end{equation}
and in the limit $\theta-z\rightarrow 0$, expand $r^{\theta-z}$ as
\begin{equation}
  r^{\theta-z}=1+(\theta-z)\ln r+\cdot\cdot\cdot,
\end{equation}
so that the recasted metric is
\begin{align}
  f=1-(m-\frac{(q_{2})^{2}}{2(\theta-2)(\theta-z)})r^{\theta-z-2}+\frac{(q_{2})^{2}r^{\theta-z-2}\ln r}{2(\theta-2)} \nonumber \\
  +\frac{B^{2}r^{2z-6}}{4(z-2)(3z-\theta-4)}-\frac{k^{2}r^{\theta-2z}}{(\theta-2)(z-2)}.
\end{align}
After defining
\begin{equation}
  M=m-\frac{(q_{2})^{2}}{2(\theta-2)(\theta-z)},
\end{equation}
we obtain
\begin{gather}
  f=1-Mr^{\theta-z-2}+\frac{(q_{2})^{2}r^{\theta-z-2}\ln r}{2(\theta-2)}+\frac{B^{2}r^{2z-6}}{4(z-2)(3z-\theta-4)}-\frac{k^{2}r^{\theta-2z}}{(\theta-2)(z-2)} \nonumber \\
  =1-\frac{M}{r^{2}}-\frac{(q_{2})^{2}\ln r}{2r^{2}}+\frac{B^{2}}{8r^{4}}-\frac{k^{2}}{r},
\end{gather}
from which the temperature is
\begin{equation}
  T=\frac{1}{4\pi}\bigg(2r_{+}-\frac{(q_{2})^{2}}{2r_{+}}-\frac{B^{2}}{4r_{+}^{3}}-k^{2}\bigg).
\end{equation}
It is consistent with (\ref{temperature}). The conductivities and thermoelectrical conductivities are straightforwardly
\begin{gather}
  \sigma_{xx}=\frac{r_{+}^{2}k^{2}(q_{2}^{2}+r_{+}^{-2}B^{2}+r_{+}k^{2})}{(r_{+}^{-1}B^{2}+r_{+}^{2}k^{2})^{2}+q_{2}^{2}B^{2}}, \nonumber \\
  \sigma_{xy}=\frac{q_{2}B(q_{2}^{2}+r_{+}^{-2}B^{2}+2r_{+}k^{2})}{(r_{+}^{-1}B^{2}+r_{+}^{2}k^{2})^{2}+q_{2}^{2}B^{2}}, \nonumber \\
  \alpha_{xx}=\frac{4\pi r_{+}^{3}q_{2}k^{2}}{(r_{+}^{-1}B^{2}+r_{+}^{2}k^{2})^{2}+q_{2}^{2}B^{2}}, \nonumber \\
  \alpha_{xy}=\frac{4\pi r_{+}B(q_{2}^{2}+r_{+}^{-2}B^{2}+r_{+}k^{2})}{(r_{+}^{-1}B^{2}+r_{+}^{2}k^{2})^{2}+q_{2}^{2}B^{2}}.
\end{gather}
The magnetoresistance, Seebeck 
coefficients are also easily to be derived
\begin{eqnarray}
\frac{\Delta\rho}{\rho}&=&\frac{r_{+}k^{2}B^{2}}{q_{2}^{2}B^{2}+r_{+}^{2}(q_{2}^{2}+r_{+}k^{2})^{2}},\\
S&=&\frac{4\pi r^3_{+} q_{2}}{r^{2}_{+}q^{2}_{2}+B^{2}+r^{3}_{+}k^{2}}.
\end{eqnarray}
And also the Hall angle is given by
\begin{equation}
\theta_{H}=\frac{q_{2}B[(r^{2}_{+}q^2_2+B^{2})+2r^{3}_{+}k^{2}]}{k^{2}r^{2}_{+}[r^{2}_{+}q^{2}_{2}+B^{2}+r^{3}_{+}k^{2}]}.
\end{equation}
The Nernst signal is defined as
\begin{equation}\label{nernst}
e_{N}\equiv (\sigma^{-1}\cdot \alpha)_{xy}=\frac{4\pi B r^{3}_{+} k^2}{B^2 q^2_2+r^{2}_{+}(q^2_2+r_{+} k^2)^2}.
\end{equation}
where $\sigma$ and $\alpha$ denote the electric and thermoelectric conductivity matrices, respectively.

\begin{figure}
\center{
\includegraphics[scale=0.4]{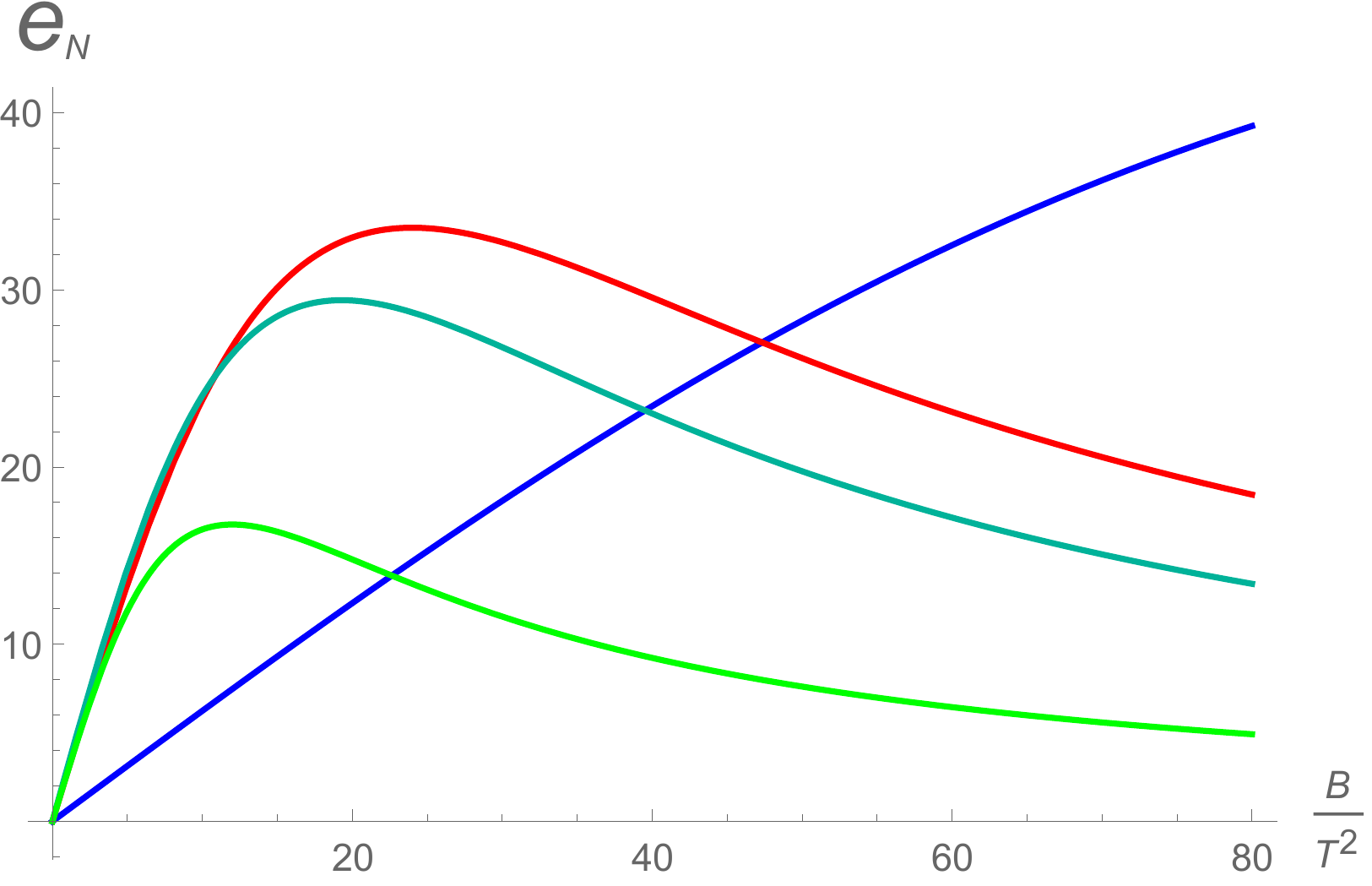}
\includegraphics[scale=0.4]{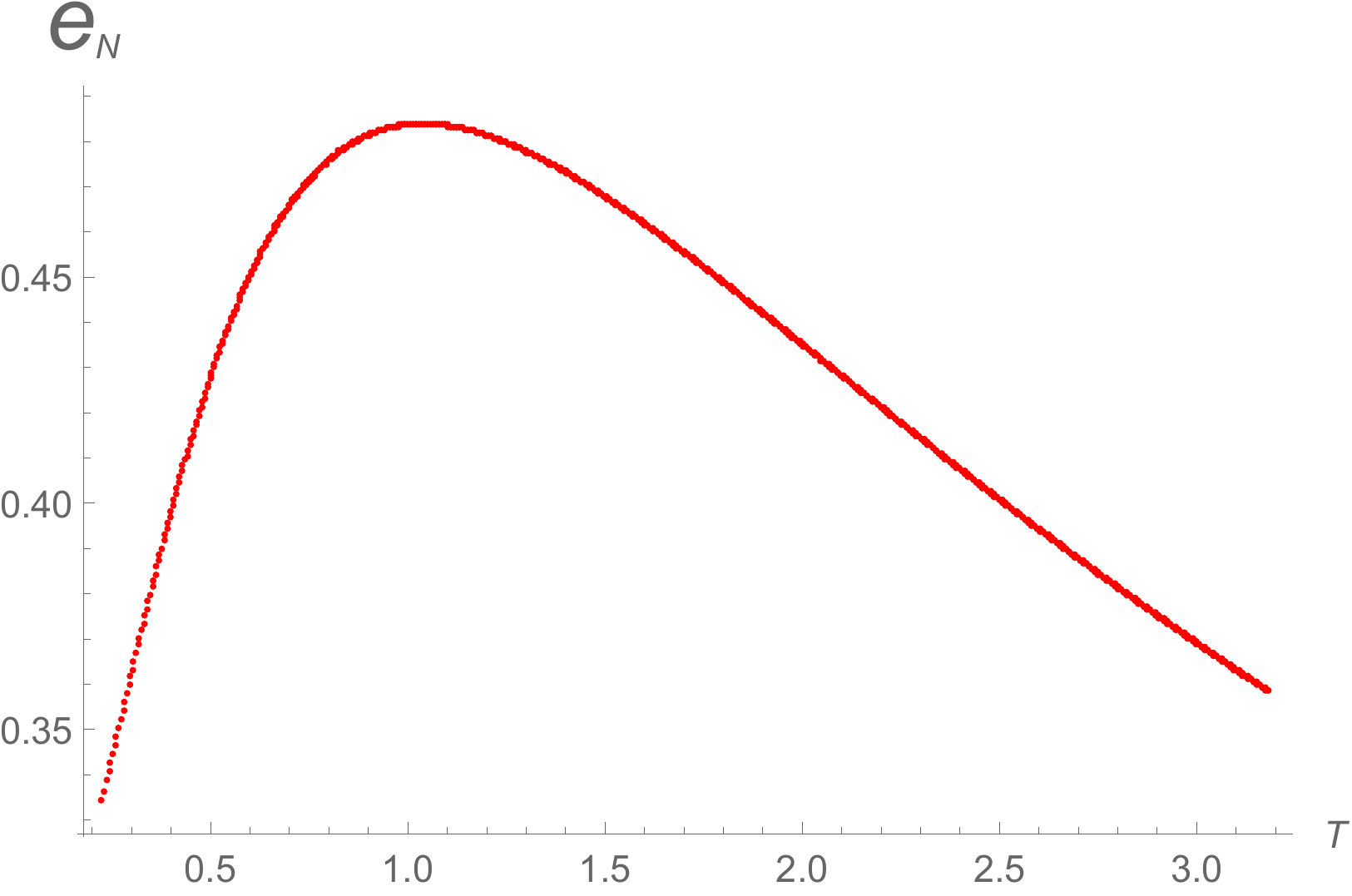}
\caption{\label{solu} (Left) Nernst signal as a function of magnetic field B. The lines from the top to bottom correspond to
$k/T=1,0.84,0.66,3$. (Right) Conventional metal behaviors Nernst signal as a function of temperature with $q=1.8$, $k=1$, and $B=1/2$. We set $\theta=1$ and $z=1$ for the first class of black hole solution.   }\label{nerst1} }
\end{figure}

In the high temperature limit, $T\sim r_{+}$, one can show that we have linear resistivity namely $\rho_{xx}\sim T$, $S$ is constant and the Hall angle satisfies the behavior $\theta_{H}\sim\frac{1}{T^{2}}$. From (\ref{nernst}) and figure \ref{nerst1}, we can see that in the high temperature limit, $e_{N}\sim \frac{4 \pi}{k^2 T}$ yields exactly bad metal behavior while in the low temperature $e_{N}\sim T$ agrees with that of conventional metals.
 These scaling behaviors are in some aspects consistent with the anomalous scaling in cuprate strange metal.

\begin{figure}
\center{
\includegraphics[scale=0.4]{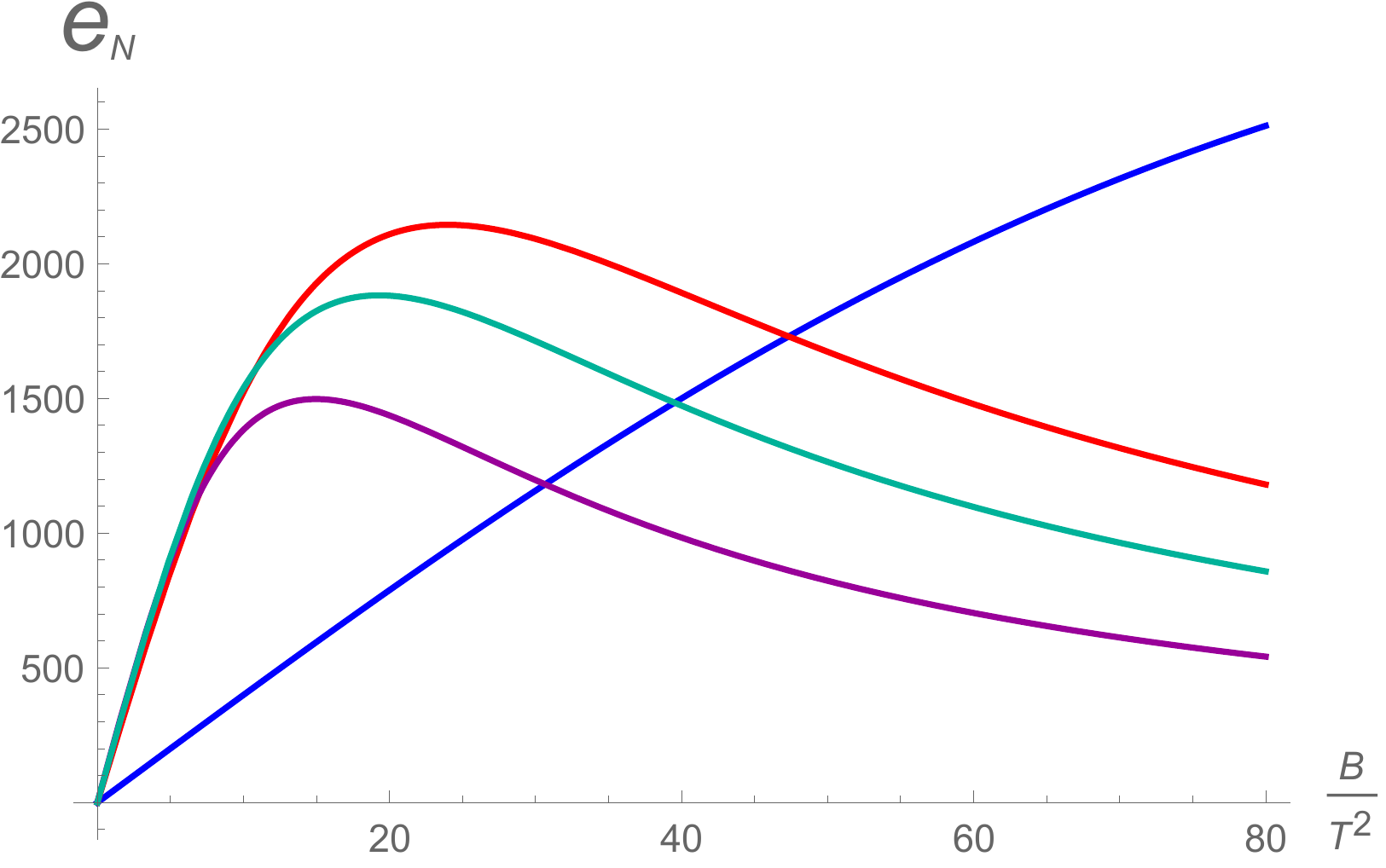}
\includegraphics[scale=0.4]{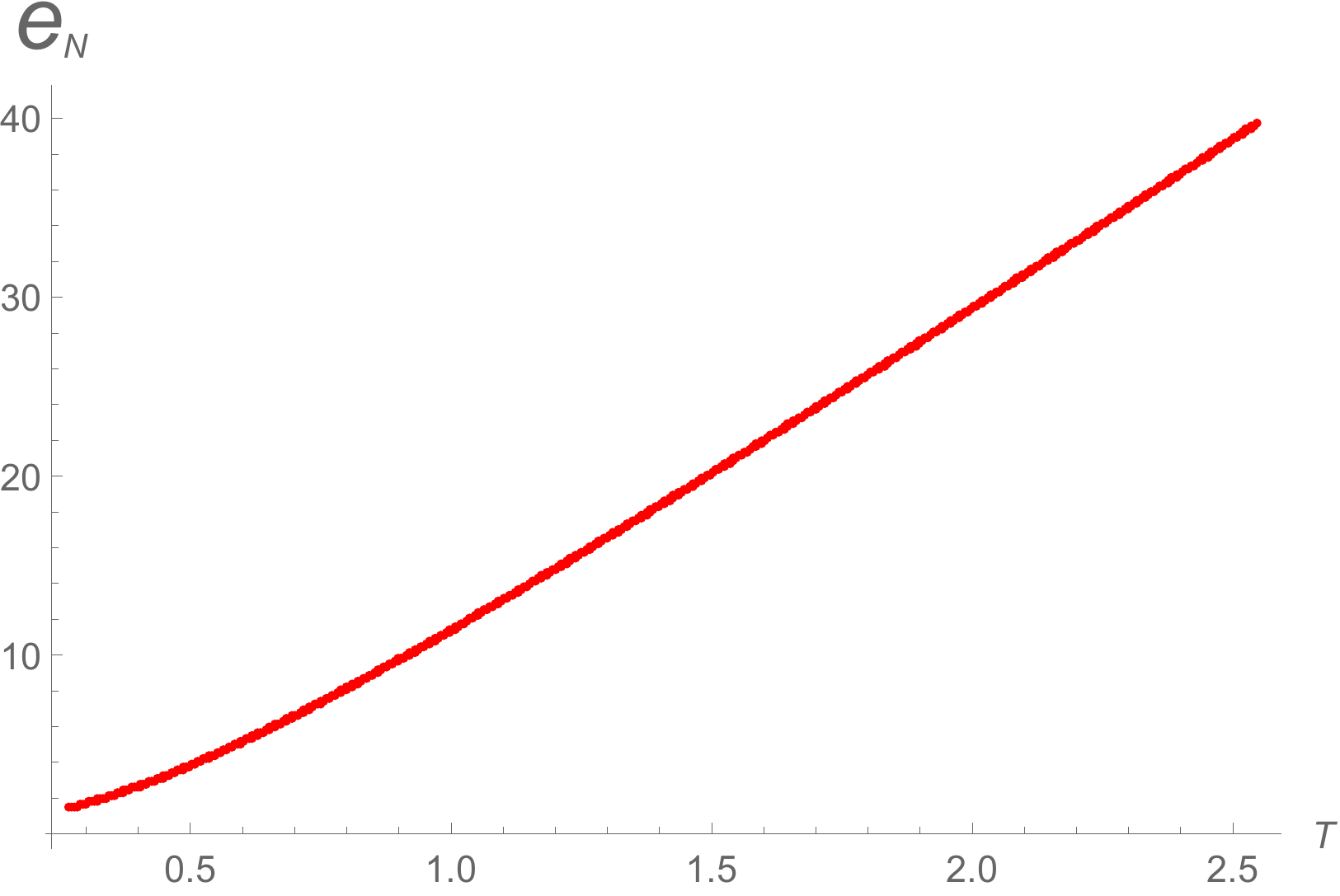}
\caption{\label{solu} (Left) Nernst signal as a function of magnetic field B. The lines from the top to bottom correspond to
$k/T=1,0.84,0.66,3$. (Right) Conventional metal behaviors Nernst signal as a function of temperature with $q=1$, $k=1$, and $B=1/2$.  We set $\theta=-1$ and $z=1$ for the second class of solution. }\label{nerst2} }
\end{figure}

\textbf{$\bullet$ Second class solution with momentum dissipation--} Next we consider another background metric (\ref{fsolution3}). (\ref{actionprameter1}) does not change except
\begin{equation}\label{actionprameter2}
  Z_{2}=r^{\theta-2z+2}.
\end{equation}
Similarly, it will return to asymptotic AdS situation when $z=1,\theta=0$. Besides, we will use the values $z=1,\theta=-1$ as a non-AdS example. After redefining the mass term
\begin{equation}
  M^{*}=m^{*}-\frac{(q_{2})^{2}}{4(z-2)(3z-\theta-4)},
\end{equation}
we can obtain the metric in an analogous way
\begin{gather}
  f^{*}=1-M^{*}r^{\theta-z-2}+\frac{(q_{2})^{2}r^{\theta-z-2}\ln r}{4(z-2)}+\frac{B^{2}r^{2\theta-2z-2}}{2(\theta-2)(\theta-z)}-\frac{k^{2}r^{\theta-2z}}{(\theta-2)(z-2)} \nonumber \\
  =1-\frac{M^{*}}{r^{4}}-\frac{(q_{2})^{2}\ln r}{4r^{4}}+\frac{B^{2}}{12r^{6}}-\frac{k^{2}}{3r^{3}}.
\end{gather}
The corresponding   temperature and conductivities can be recast as
\begin{gather}
  T=\frac{1}{4\pi}(4r_{+}-\frac{(q_{2})^{2}}{4r_{+}^{3}}-\frac{B^{2}}{6r_{+}^{5}}-\frac{k^{2}}{3r_{+}^{2}}), \nonumber \\
  \sigma_{xx}=\frac{r_{+}^{2}k^{2}(q_{2}^{2}+r_{+}^{-2}B^{2}+r_{+}k^{2})}{(r_{+}^{-1}B^{2}+r_{+}^{2}k^{2})^{2}+q_{2}^{2}B^{2}}, \nonumber \\
  \sigma_{xy}=\frac{q_{2}B(q_{2}^{2}+r_{+}^{-2}B^{2}+2r_{+}k^{2})}{(r_{+}^{-1}B^{2}+r_{+}^{2}k^{2})^{2}+q_{2}^{2}B^{2}}, \nonumber \\
  \alpha_{xx}=\frac{4\pi r_{+}^{5}q_{2}k^{2}}{(r_{+}^{-1}B^{2}+r_{+}^{2}k^{2})^{2}+q_{2}^{2}B^{2}}, \nonumber \\
  \alpha_{xy}=\frac{4\pi r_{+}^{3}B(q_{2}^{2}+r_{+}^{-2}B^{2}+r_{+}k^{2})}{(r_{+}^{-1}B^{2}+r_{+}^{2}k^{2})^{2}+q_{2}^{2}B^{2}}.
\end{gather}
We can see the electric conductivities are the same as those in the case of former background metric when $z=1,\theta=1$, while the thermoelectric conductivities are not.
Similarly, it is also straightforward to derive the Seebeck coefficient, magnetoresistance and Hall angle for this case. Then we have the Seebeck coefficient
\begin{equation}
S=\frac{4\pi q_{2}r^{3}_{+}}{q^{2}_{2}+r^{-2}_{+}B^{2}+r_{+}k^{2}},
\end{equation}
the magnetoresistance
\begin{equation}
\frac{\Delta\rho}{\rho}=\frac{r_{+}k^{2}B^{2}}{q_{2}^{2}B^{2}+r_{+}^{2}(q_{2}^{2}+r_{+}k^{2})^{2}},
\end{equation}
and the Hall angle
\begin{equation}
\theta_{H}=\frac{q_{2}B(q^{2}_{2}+r^{-2}_{+}B^{2}+2r_{+}k^{2})}{r^{2}_{+}k^{2}(q^{2}_{2}+r^{-2}_{+}B^{2}+r_{+}k^{2})}.
\end{equation}
The Nernst signal can also be evaluated
\begin{equation}
e_{N}=\frac{4\pi B r^{5}_{+} k^2}{B^2 q^2_2+r^{2}_{+}(q^2_2+r_{+} k^2)^2}.
\end{equation}
Again, in high temperature limit, we have the following scaling behavior, $\rho_{xx}\sim T$, $\theta_{H}\sim\frac{1}{T^{2}}$, $S\sim T^{2}$, $\frac{\Delta\rho}{\rho}\sim \frac{B^{2}}{T^{3}}$ and $e_{N}\sim T$. Although we have linear resistivity and quadratic temperature dependence of inverse Hall angle as the cuprate strange metal, but the temperature dependence of Seebeck coefficient and magnetoresistance do not match with the scaling behavior found in cuprate experiments. Meanwhile, the Nernst signal also shows its conventional metal behavior as shown in figure \ref{nerst2}.

    \textbf{$\bullet$ Two classes  without momentum dissipation--} Another interesting case is when $k=0$, but keeping $q_1$ and $q_2$ finite. It was noticed in \cite{Ge:2016sel} that in this case the linear resistivity can also be realized even without momentum dissipation.  The resulting thermoelectric conductivities become
    \bea
  \sigma_{xx}&=& \sigma_{yy} =\frac{q_{1}^{2}[Z_{1}(q_{2}^{2}+Z_{2}^{2}B^{2})+Z_{2}(q_{1}^{2})]}
             {[q_{1}^{2}+Z_{1}(Z_{2}B^{2})]^{2}+Z_{1}^{2}q_{2}^{2}B^{2}}, \nonumber \\
  \sigma_{xy}&=&-\sigma_{yx}=\frac{Z_{1}q_{2}B[Z_{1}(q_{2}^{2}+Z_{2}^{2}B^{2})+2Z_{2}(q_{1}^{2})]}
             {[q_{1}^{2}+Z_{1}(Z_{2}B^{2})]^{2}+Z_{1}^{2}q_{2}^{2}B^{2}}, \nonumber \\
  \alpha_{xx}&=& \alpha_{yy}=\frac{4\pi Z_{1}Wq_{2}q_{1}^{2}}
             {[q_{1}^{2}+Z_{1}Z_{2}B^{2}]^{2}+Z_{1}^{2}q_{2}^{2}B^{2}}, \nonumber \\
  \alpha_{xy}&=&-\alpha_{yx}=\frac{Z_{1}q_{2}B[Z_{1}(q_{2}^{2}+Z_{2}^{2}B^{2})+2Z_{2}q_{1}^{2}]}
             {[q_{1}^{2}+Z_{1}Z_{2}B^{2}]^{2}+Z_{1}^{2}q_{2}^{2}B^{2}}.
\eea
In the absence of magnetic field, namely $B=0$, the electrical conductivities reduces to
\bea
\sigma_{xx}= \sigma_{yy}=Z_2+\frac{q^2_2 Z_1}{q^2_1}, \sigma_{xy}= \sigma_{yx}= 0.
\eea
Assuming the first term in $\sigma_{xx}$ corresponds to  the intrinsic current relaxation effect and thus leads to the linear in temperature resistivity, we obtain $\theta=8/5$ and $z=6/5$ for the first class of the black hole solution so that both linear and quadratic in temperature resistivities can be realized. On the other hand, we need $\theta=0$ and $z=2$ for the second class of solution to realize the linear and quadratic in temperature resistivities.
The Seebeck coefficient is then given by
\be
S=\frac{4\pi Z_1 W q_2 q^2_1}{q^2_1(Z_1 q^2_2+Z_2 q^2_1)}
\ee
For the first class of the black hole solution with $\theta=8/5$ and $z=6/5$, the Seebeck coefficient scales as $S\sim T^{-2/3}$ in the high temperature limit. But for the second class of the black hole solution, we have $S \sim constant $ in the high temperature limit.

In the presence of magnetic field, we are able to calculate the magnetoresistance. For both classes of the solution, we have $\Delta\rho/\rho\sim B^2/T^3$ for both cases ($\theta=8/5$ , $z=6/5$) and ($\theta=0$, $z=2$). The Nernst signal for the first class of solution with ($\theta=8/5$ , $z=6/5$) is given by
\be
e_{N}=\frac{4\pi B q^2_1 r_{+}^{14/5}}{B^2 q^2_2+r_{+}^{12/5}(q^2_2+q^2_1 r_{+}^{6/5})^2}.
\ee
\begin{figure}
\center{
\includegraphics[scale=0.4]{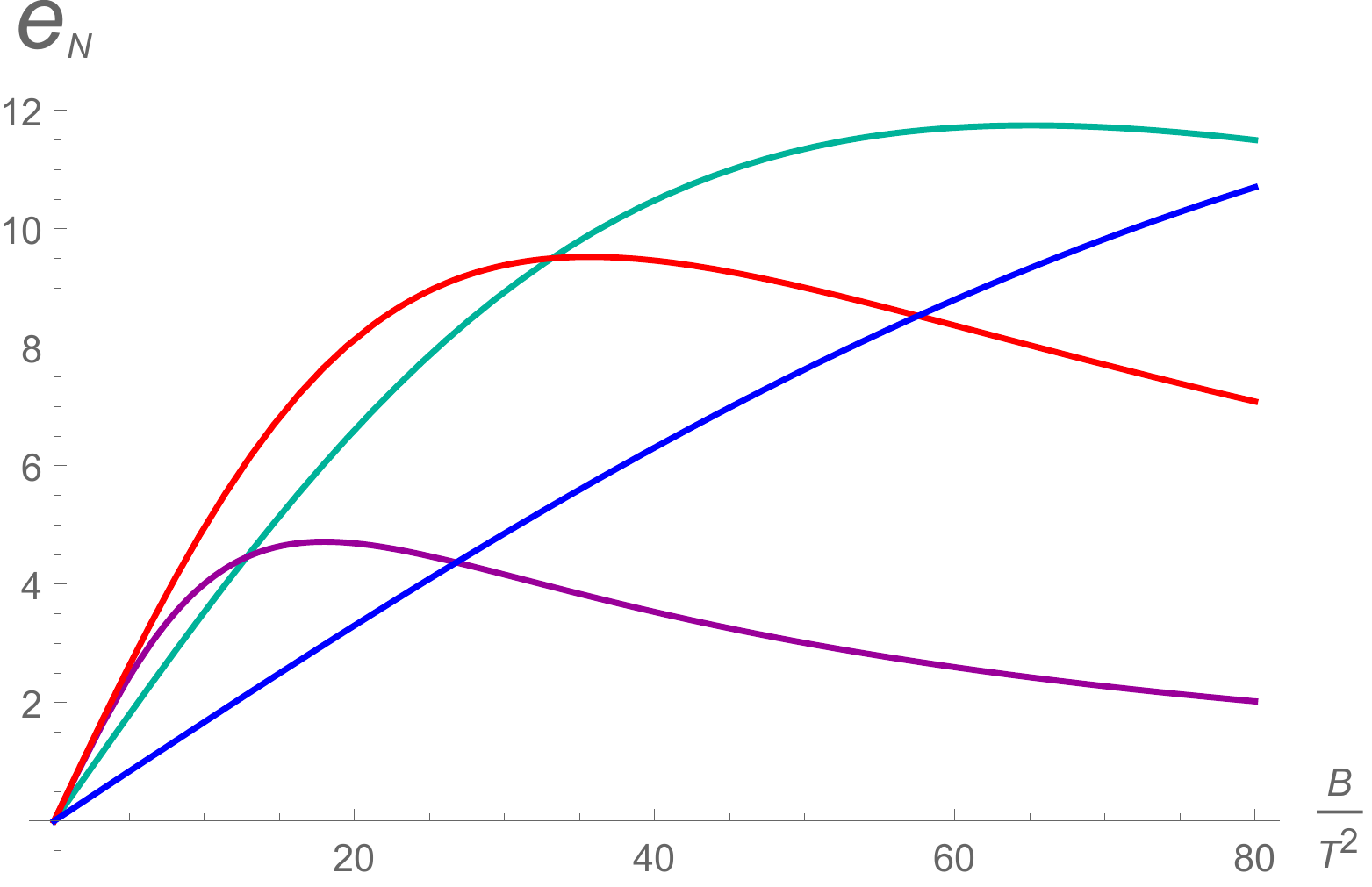}
\includegraphics[scale=0.4]{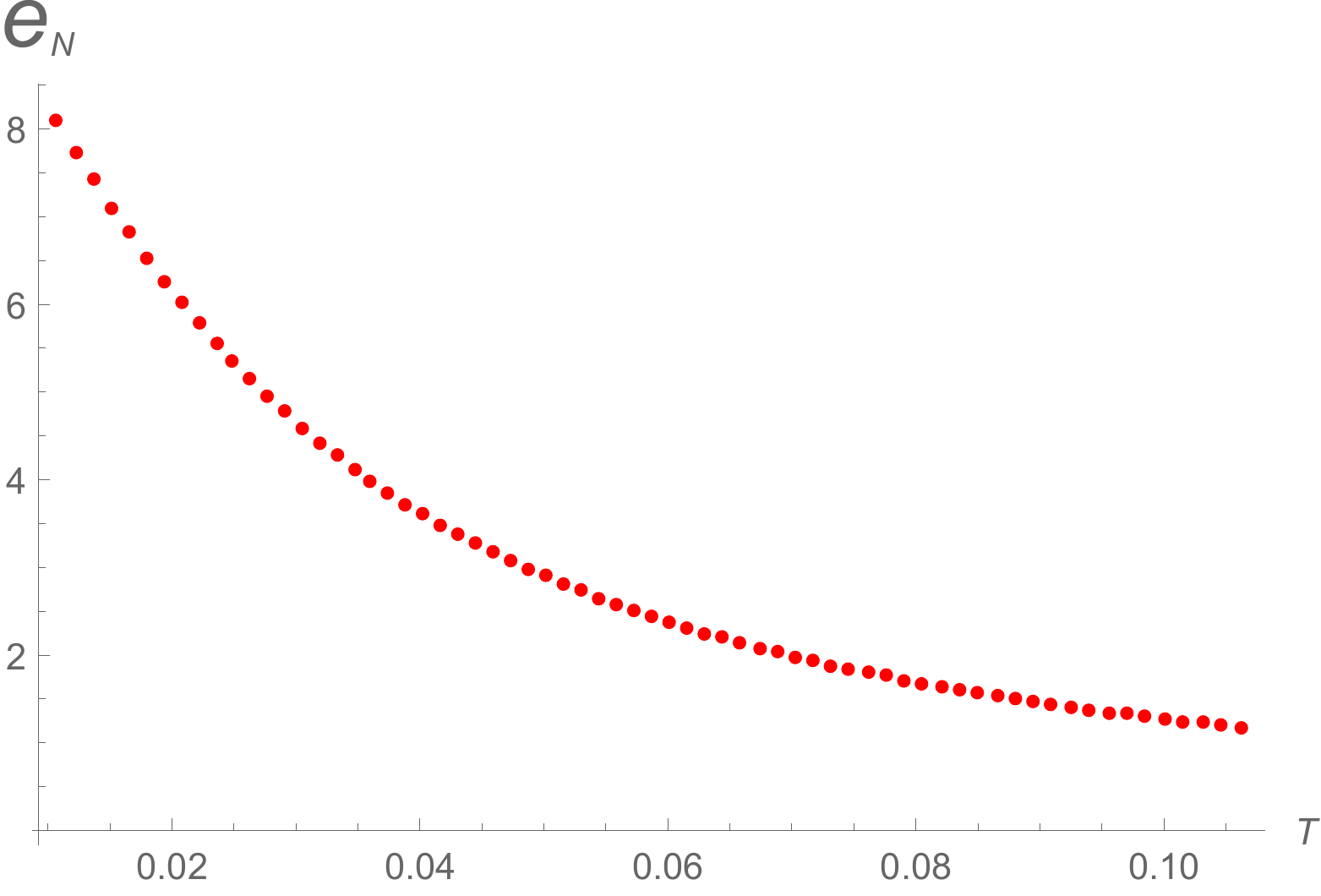}
\caption{ (Left) Nernst signal as a function of magnetic field B. The lines from  right to left correspond to
$q_1/T=0.2,0.4.0.6,1$. (Right)  Non-conventional metal behavior of the Nernst signal as a function of temperature with $q_1=1$, $q_2=1/5$  and $B=1/10$. We consider the first class of the black hole solution and  set $\theta=8/5$ and $z=6/5$.   } \label{nerst30}}
\end{figure}
From figure \ref{nerst30}, we can see that the magnetic dependence of the Nernst signal is quiet similar to the previous cases. But the temperature-dependence of the signal only shows its bad-mental-like behaviors as $e_{N}\sim 1/T^{5/3}$ in the high temperature limit.

\begin{figure}
\center{
\includegraphics[scale=0.4]{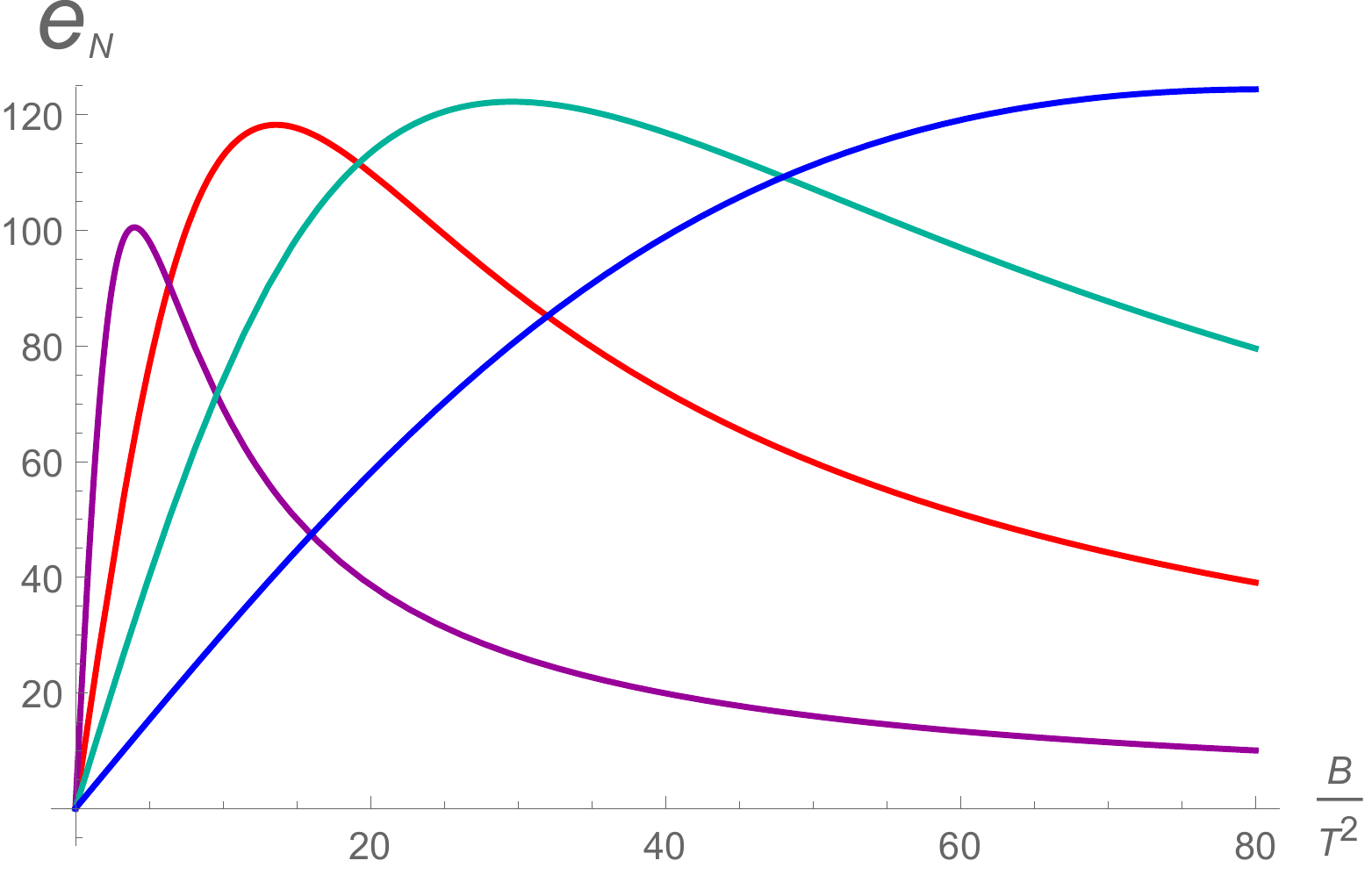}
\includegraphics[scale=0.4]{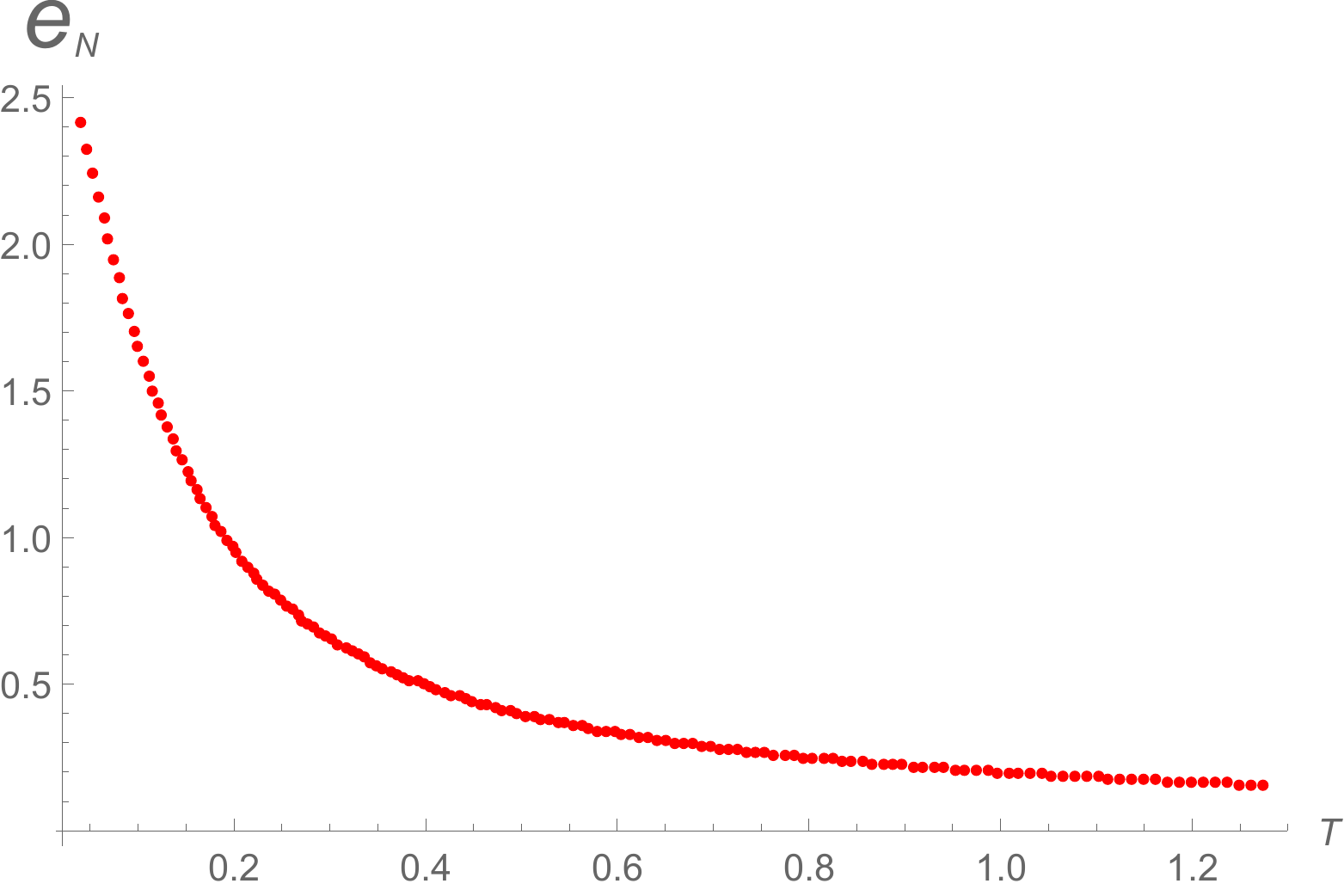}
\caption{(Left) Nernst signal as a function of magnetic field B. The lines from  right to left correspond to
$q_1/T=0.2,0.4.0.6,1$. (Right) Non-conventional metal behavior of the Nernst signal as a function of temperature with $q_1=1$, $q_2=1/5$  and $B=1/10$. We consider the second class of the black hole solution and  set $\theta=0$ and $z=2$.  } \label{nerst4}}
\end{figure}

For the second class of black hole solution with $\theta=0$, $z=2$, the Nernst signal goes as
\be
e_{N}=\frac{4\pi B q^2_1 r_{+}^{6}}{B^2 q^2_2+r_{+}^{4}(q^2_2+q^2_1 r_{+}^{2})^2}.
\ee
As shown in figure \ref{nerst4}, the Nernst signal scales as $e_{N}\sim 1/T$ in the high temperature regime. But in the low temperature, the conventional metal behavior $e_{N}\sim T$ cannot be recovered.  The magnetic dependence of the Nernst signal is more close to the experimental result found in \cite{wang}.
\section{Conclusion}

In this paper, we studied the Einstein-Maxwell-Dilaton model with massless Axion fields providing momentum dissipation and obtain two classes of the analytical black hole solutions in the presence of an external magnet field when spacetime yields Lifshitz scaling. The effect of the hyperscaling factor was also considered.  The magnetothermoelectric DC conductivities were thus calculated in terms of horizon data by means of holographic principle.

In order to mimic the experimental results, we consider special choices of the values of the dynamical and the hyperscaling factors. For the first class of black hole solution with momentum dissipation, we found that $\theta=1$ and $z=1$ could leads to linear and quadratic in temperature resistivities, inverse T square Hall angle and experiment compatible  Seebeck coefficient.  Remarkably, the Nernst signal shows  exactly bad metal behaviors in the high temperature regime (i.e. $e_{N}\sim 1/T $) and conventional metal behavior $e_{N}\sim T$ in the low temperature region, in good agreement with the experimental results of cuprates. Although the null energy condition is violated, a careful check of the local thermodynamic stability and the casual structure of the boundary theory reveals that it is locally stable and free from superluminal signal propagation on the
boundary.  For the second class of black hole solution with momentum dissipation with $\theta=-1$ and $z=1$, the linear and quadratic in temperature resistivities can still be realized. But the Seebeck coefficient shows $S\sim T^2$ scaling and the Nernst signal only yields conventional metal behavior.

As a byproduct of this paper, we realized that even in the absence of momentum dissipation,  the DC electrical conductivity still has two terms of contributions and the dual conductivity is finite. We discuss in detail how to reproduce the anomalous transport of cuprates for these two classes of black hole solution. For the first class, $\theta=8/5$ and $z=6/5$ results in linear and quadratic in temperature resistivities, leaving the Seebeck coefficient different from the experiments and non-conventional metal behavior of the Nernst signal.  For the second class, $\theta=0$ and $z=2$, linear and quadratic in temperature resistivities can be realized without trouble. The Seebeck coefficient scales as $S\sim constant$. But the Nernst signal only marks non-conventional metal behaviors.
\section*{Acknowledgements}

This work was partly supported by NSFC, China (No.11375110) and No. 14DZ2260700 from Shanghai Key Laboratory of High Temperature Superconductors. SYW was supported by the Minister of Science and Technology (grant no. 105-2112-M-009-001-MY3) in Taiwan.

\end{document}